\documentclass{mem}
\usepackage{natbib}\usepackage{txfonts}\usepackage{balance}
\usepackage{graphicx}
\usepackage[a4paper,breaklinks,dvipdfm]{hyperref}
\idline{0}{0}
\begin{document}
\def\teff{$T\rm_{eff }$}
\def\kms{$\mathrm {km s}^{-1}$}

\title{The Metallicity of High-$z$ GRBs: the Case of GRB 111008A}

   \subtitle{}

\author{
Martin Sparre
          }

  \offprints{Martin Sparre}

\institute{
Dark Cosmology Centre,
Niels Bohr Institute, University of Copenhagen\\
Juliane Maries Vej 30, 2100 Copenhagen O, Denmark\\
\email{sparre@dark-cosmology.dk}
}

\authorrunning{Sparre}

\titlerunning{The Metallicity of High-$z$ GRBs}

\abstract{
With optical and near-infrared follow-up observations of gamma-ray bursts (GRBs) it is possible to study the chemical enrichment of galaxies at high redshift. Especially interesting are measurements of the metallicity of GRB host with $z\gtrsim 4.7$, since it has recently been shown that damped Lymann-$\alpha$ absorbers have a rapid decrease in metallicity at this redshift \citep{Rafelski2013}. In this paper measurements of the amount of metals and dust in the host galaxy of GRB 111008A at $z=5.0$ is presented, and it is discussed how observations of more high-$z$ GRB host galaxies can give clues about dust formation mechanisms and the distribution of metals in the early Universe.

}
\maketitle{}

\section{Motivation}

Long gamma-ray bursts are lighthouses that can be used to study the line of sight towards nearby and distant galaxies. Signatures from a core collapse supernova have been observed in the optical spectra of several long GRBs (e.g. \citealt{Sparre2011,Hjorth2003}). An implication is that GRBs originate in galaxies with recent star formation, but there is no consensus about whether GRBs are equally likely to be produced by stars of all metallicities. Theoretically, it is predicted that a collapsing star is more likely to produce a GRB if it has a low metallicity \citep{WH2006}, since low-metallicity stars typically have a higher angular momentum, which is important for creating a GRB jet. Observations of high-metallicity GRB host galaxies do, however, challenge this picture \citep[e.g. GRB 020819,][]{Levesque2010}.

A fraction of GRBs have host galaxies that gives rise to damped Lyman-$\alpha$ absorption; these systems are called GRB-DLAs \citep{Jakobsson2006,Fynbo2009}. GRB-DLAs are selected in a different way than DLAs (QSO-DLAs), which happens to be along the sightlines to distant quasars; the latter are selected by the HI-cross sections with
$\log N(\textrm{HI})/\textrm{cm}^{-2}>20.3$
 for photons traveling randomly through the Universe, and the former probes photon sightlines that origin in the deep interior of star-forming galaxies \citep{Prochaska2008}. Statistically GRB-DLAs have a metallicity which is in the upper end of the distribution seen for QSO-DLAs. The metallicity distributions are expected to be different due to the different way their sightlines are selected \citep{Fynbo2008}.

In this proceeding it is discussed how the observations of high-$z$ gamma-ray bursts can be used to increase the understanding of the metallicity of the neutral gas at high redshift. As an example I present the amount of dust and metals in GRB 111008A at $z=5.0$.

\begin{table*}
\caption{The abundance pattern of the elements (including excited levels) in the host galaxy of GRB 111008A measured with the VLT/X-shooter spectrograph.}
\label{table:metallicity}
\centering 
\begin{tabular}{lcc}
\hline\hline 
Element & $\log N(X) / $cm$^{-2}$ & $[$X$/$H$]$\\ \hline 
H&$22.30\pm0.06$&-\\
S & $15.71 \pm 0.09 $ & $ -1.70 \pm 0.10$\\
Cr & $14.17 \pm 0.09 $ &$ -1.76 \pm 0.11$\\
Fe & $16.05 \pm 0.05 $ & $-1.74 \pm 0.08$\\
Ni&$14.89 \pm 0.18 $ & $-1.64 \pm 0.19$\\
\hline\hline
\end{tabular}
\end{table*}

\section{The Metallicity of the Host Galaxy of GRB 111008A at $z=5.0$}

\subsection{The Chemical Composition}

From VLT/X-shooter spectroscopy of the optical afterglow of GRB 111008A at $z=5.0$ \citep{Sparre2013}, it is determined that this bursts host galaxy had an HI-content of
\begin{eqnarray}
\log N(\textrm{HI})/\textrm{cm}^{-2}=22.30 \pm 0.06,
\end{eqnarray}
and a metallicity content of 
\begin{eqnarray}
[\textrm{S}/\textrm{H}]=-1.70 \pm 0.10.
\end{eqnarray}
GRB 111008A is currently the highest redshift GRB host galaxy with such a precise measurement of the metallicity. A summary of the elements with reliably determined abundances are in Table~\ref{table:metallicity}.

\subsection{Dust-to-metals Ratio and an Intervening DLA at $z=4.6$}

By combining data from the instruments, SWIFT/XRT and GROND, a SED was constructed, and the $V$-band extinction from dust towards the sightline was fitted to be $A_V=0.11 \pm 0.04$. The dust-to-metals ratio of the gas in GRB 111008As host galaxy was constrained to be equal to or lower than what is observed for the gas in the Local Group.

The sightline to the GRB 111008A host galaxy was affected by an intervening DLA with $\log N(\textrm{HI})/\textrm{cm}^{-2}=21.34 \pm 0.10$ at $z=4.6$. The measured $A_V$ along the sightline is therefore the sum of the extinction for both this intervening DLA and the GRB host galaxy DLA.

If the intervening absorber had not been present it would of course have been possible to do derive a stronger constraint on the $A_V$ and the dust-to-metals ratio of the GRB 111008A host galaxy.

\section{Why are Observations of more $z\approx 5$ GRBs Important?}

The study of GRB 111008A shows that long GRBs provide a good tool for studying the high-$z$ Universe. Observations of more GRBs at a similar redshift of $z\sim 5$ is important for e.g. the following reasons:

\begin{itemize}
\item Recently, \citet{Rafelski2013} shows that the metallicity of QSO-DLAs decrease rapidly at $z \gtrsim 4.7$ compared to the distribution below this redshift. Observations of more GRB-DLAs (e.g. 5-10 systems) at $z\gtrsim 4.7$ are needed to show whether a similar trend is present in the abundances of GRB-DLAs. It is not necessarily expected that GRB-DLAs will also have such a rapid decrease in metallicity at $z \gtrsim 4.7$, since GRB-DLAs are selected in a different way than QSO-DLAs.
\item It is unknown how the dust is formed in the early Universe; some of the possible mechanisms are production in supernovae \citep{Morgan2003} and interstellar grain growth \citep{Draine2009}. If dust is formed in supernovae it is expected that the dust-to-metals ratio of galaxies are independent of the metallicity, whereas a lower dust-to-metals ratio is expected at low metallicity in the scenario of growth of interstellar dust grains. With GRB 111008A we could constrain the dust-to-metals ratio as well as the metallicity, but the error on the dust-to-metals ratio is large, since it is unclear how large a fraction of the $A_V$ that was due the intervening absorber at $z=4.6$. With a bit of luck, observations of similar GRB hosts without a strong intervening absorption system will help to further constrain the dust production mechanism in the early Universe.
\end{itemize}

Finally we note that several other high-$z$ GRBs have well-constrained metallicities, e.g.

\begin{itemize}
\item \citet{Chornock2013} constrains the metallicity of GRB 130606A at $z = 5.913$ to be $[$Si$/$H$] \gtrsim -1.7$ and  $[$S$/$H$] \lesssim -0.5$.
\item \citet{Thone2013} observed GRB 100219A at $z=4.7$ and measured a metallicity of $[$S$/$H$] = -1.1 \pm 0.2$. The abundance pattern is affected by either strong dust depletion or over-abundance of $\alpha$-elements.
\end{itemize}

\begin{acknowledgements}
Thanks to Johan P. U. Fynbo for useful comments.
\end{acknowledgements}

\bibliographystyle{aa}

\end{document}